\documentclass[a4paper,fleqn,usenatbib]{mnras}

\usepackage{latexsym}
\usepackage{amsmath}
\usepackage{amssymb}
\usepackage{fnpos}
\usepackage{hyperref}
\usepackage{tabularx}
\usepackage{xspace}
\usepackage{enumerate}
\usepackage{scalerel}
\usepackage{graphicx}
\usepackage{url}
\usepackage{caption}
\usepackage{gensymb}
\usepackage{longtable}
\usepackage{lscape}
\usepackage[normalem]{ulem} 
\usepackage[dvipsnames]{xcolor} 
\usepackage{wasysym} 
\definecolor{darkgreen}{rgb}{0.0,0.55,0.0}
\definecolor{darkblue}{rgb}{0.0,0.0,0.5}

\newcommand{\eal}[2]{\ifmmode{\mathrm{#1\,#2}}\else{#1\textsc{$\,$\lowercase{#2}}}\fi\xspace}
\newcommand{\feal}[2]{\ifmmode{\mathrm{#1\,#2}}\else{[#1\textsc{$\,$\lowercase{#2}}]}\fi\xspace}
\newcommand{\hfeal}[2]{\ifmmode{\mathrm{#1\,#2}}\else{#1\textsc{$\,$\lowercase{#2}}]}\fi\xspace}

\newcommand{\orcid}[1]{$^{\rm \href{https://orcid.org/#1}{\includegraphics[height=0.6em]{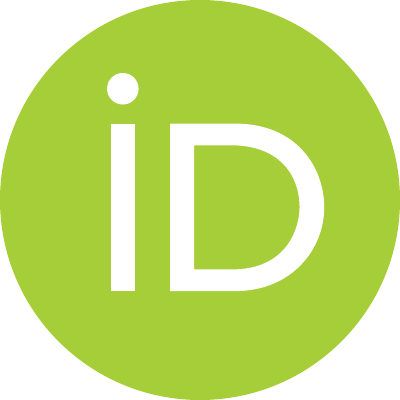}}}$}

\def\jaavso{JAAVSO}

\title[The AM CVn system ASASSN-21br]{Spectro-photometric follow up of the outbursting AM CVn system ASASSN-21br}
\author[Painter et al.]{\parbox{\textwidth}{S.~Painter$^{1}$, E.~Aydi\orcid{0000-0001-8525-3442}$^{1}$\thanks{Elias Aydi - NFHP Hubble Fellow; E-mail: aydielia@msu.edu}, M.~Motsoaledi$^{2,3}$ K.~V.~Sokolovsky\orcid{0000-0001-5991-6863}$^{4, 5}$, J.~Strader$^{1}$, D.~A.~H.~Buckley$^{2, 3}$, 
C.~S.~Kochanek$^{6}$ T.~J.~Maccarone$^{7}$, K.~Mukai$^{8,9}$, B.~J.~Shappee$^{10}$, and K.~Z.~Stanek$^{6}$}
\vspace{0.4cm}\\
\parbox{\textwidth}{
$^{1}$Center for Data Intensive and Time Domain Astronomy, Department of Physics and Astronomy, Michigan State University, East Lansing, MI 48824, USA\\
$^{2}$South African Astronomical Observatory, P.O.\ Box 9, 7935 Observatory, South Africa\\
$^{3}$Department of Astronomy, University of Cape Town, Private Bag X3, Rondebosch 7701, South Africa\\
$^{4}$Department of Astronomy, University of Illinois at Urbana-Champaign, 1002 W. Green Street, Urbana, IL 61801, USA\\
$^{5}$Sternberg Astronomical Institute, Moscow State University, Universitetskii~pr.~13, 119992~Moscow, Russia\\
$^{6}$Department of Astronomy, The Ohio State University, 140 West 18th Avenue, Columbus, OH 43210, USA\\
$^{7}$Department of Physics \& Astronomy, Texas Tech University, Box 41051, Lubbock, TX, 79409-1051, USA\\
$^{8}$CRESST and X-ray Astrophysics Laboratory, NASA/GSFC, Greenbelt, MD 20771, USA\\
$^{9}$Department of Physics, University of Maryland, Baltimore County, 1000 Hilltop Circle, Baltimore, MD 21250, USA\\
$^{10}$Institute for Astronomy, University of Hawaii, 2680 Woodlawn Drive, Honolulu, HI 96822, USA\\
}}

\pubyear{2024}

\begin{document}
\label{firstpage}
\pagerange{\pageref{firstpage}--\pageref{lastpage}}
\maketitle

\begin{abstract}
We report on spectroscopic and photometric observations of the AM CVn system ASASSN-21br, which was discovered in outburst by the All-Sky Automated Survey for Supernovae in February 2021. The outburst lasted for around three weeks, and exhibited a pronounced brightness dip for $\approx$ 4 days, during which the spectra showed a sudden transition from emission- to absorption-line dominated. 
Only $\approx$ 60 AM CVn systems with derived orbital periods are found in the Galaxy, therefore increasing the sample of AM CVn systems with known orbital periods is of tremendous importance to (1) constrain the physical mechanisms of their outbursts and (2) establish a better understanding of the low-frequency background noise of future gravitational wave surveys. Time-resolved photometry taken during the outburst of ASASSN-21br showed modulation with a period of around 36.65 minutes, which is likely the superhump or orbital period of the system. Time-resolved spectroscopy taken with the Southern African Large Telescope did not show any sign of periodicity in the He I absorption lines. This is possibly due to the origin of these lines in the outbursting accretion disc, which makes it challenging to retrieve periodicity from the spectral lines. 
Future follow up spectral observations during quiescence might allow us better constrain the orbital period of ASASSN-21br. 
\end{abstract}

\begin{keywords}
stars: AM CVn, cataclysmic variables --- white dwarfs.
\end{keywords}

\section{Introduction}

AM Canum Venaticorum (AM CVn) are 
compact mass-transferring binary stars, named after their prototypical star system.
The primary star in these systems is a white dwarf (WD) accreting material from a semi-degenerate helium star or another WD, which has filled its Roche-lobe (see \citealt{Warner_1995,Solheim_2010} for a review). AM CVn systems are identified by the presence of He lines and absence of hydrogen lines in their spectra. They are also characterized by short orbital periods, ranging between 5 and 70 minutes \citep{Solheim_2010,Levitan_etal_2015,Ramsay_etal_2018}, making them sources of low-frequency gravitational waves, which will be detected by future space-based gravitational waves experiments such as LISA \citep{Stroeer_Vecchio_2006,Kupfer_etal_2018}. AM CVn systems are also potential progenitors of Type Ia supernovae, and serve as probes for the final stages of low-mass binary evolution \citep{Bildsten_etal_2007}. 

\begin{figure*}
    \centering
    \includegraphics[width=\textwidth]{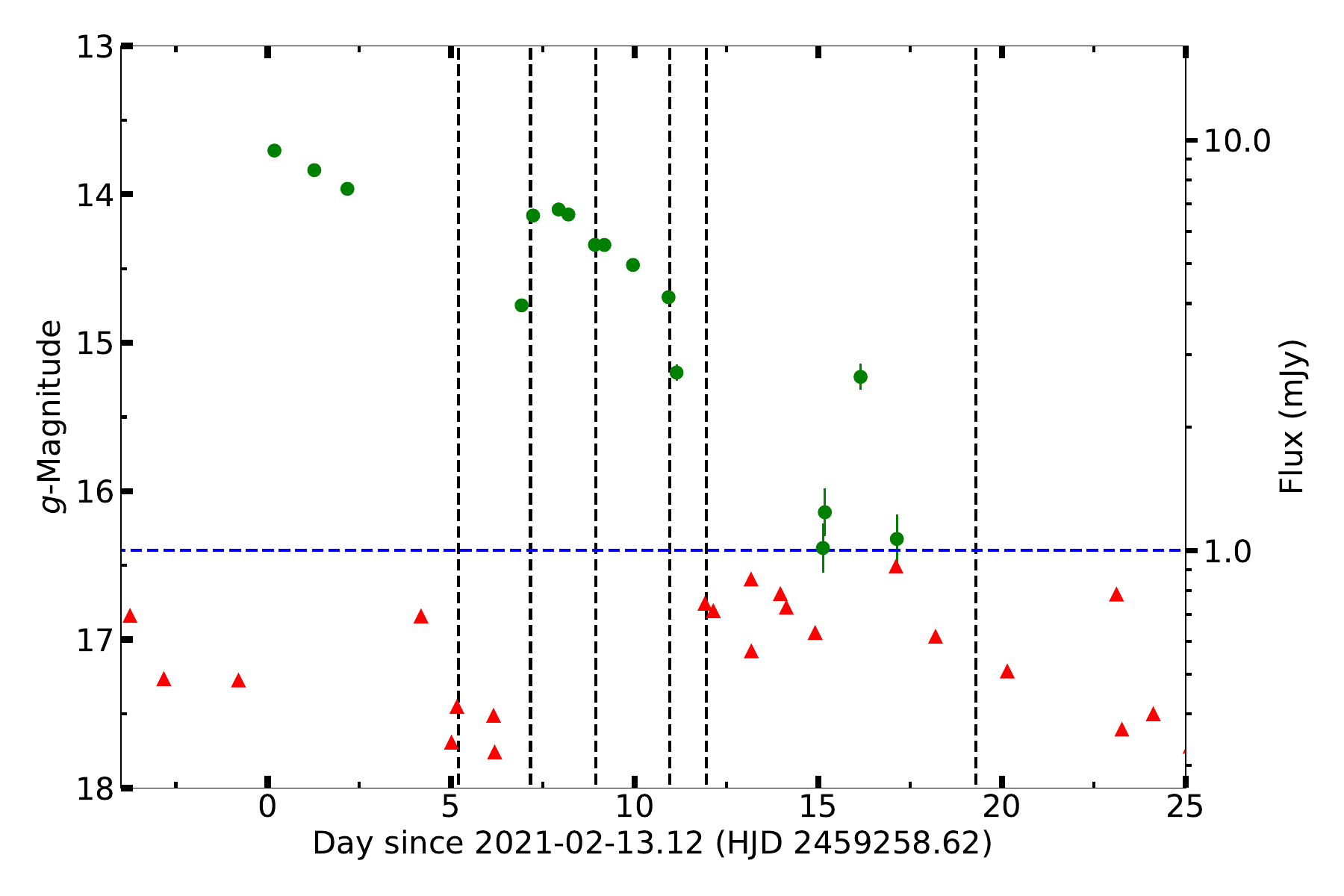}
    \caption{The ASAS-SN light curve of ASASSN-21br taken during outburst. The black vertical dashed lines mark the spectroscopic epochs. Measurement below the blue dashed line (red triangles), characterized by a flux lower than 1 mJy, are considered non-detection.}
    \label{fig:ASASSN_LC}
\end{figure*}

\begin{figure}
    \centering
    \includegraphics[width=\columnwidth]{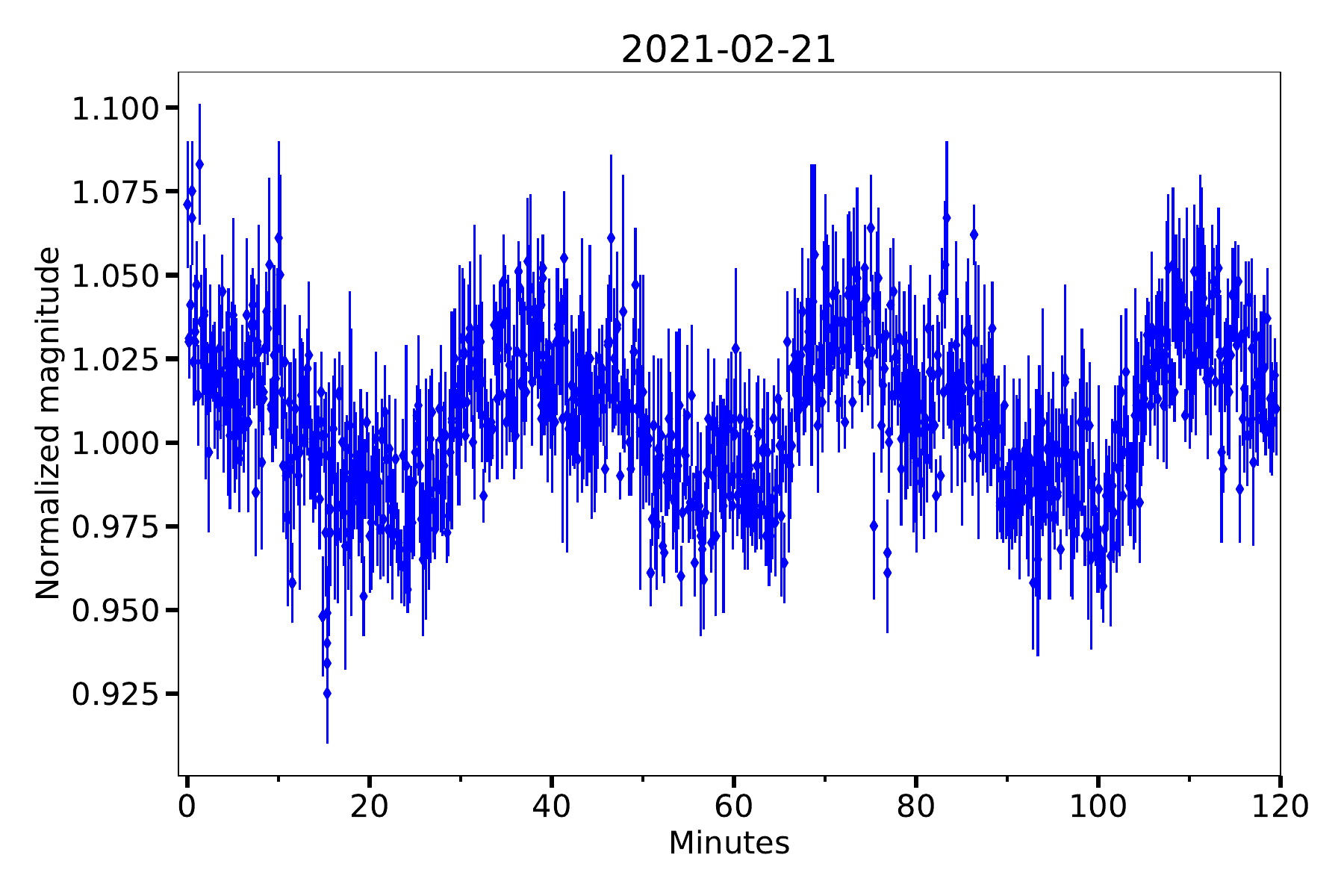}
    \caption{A sample of the SHOC photometry taken on 2021-02-21, with a cadence of 10s.}
    \label{fig:Photometry}
\end{figure}

\begin{figure*}
    \centering
    \includegraphics[width=0.49\textwidth]{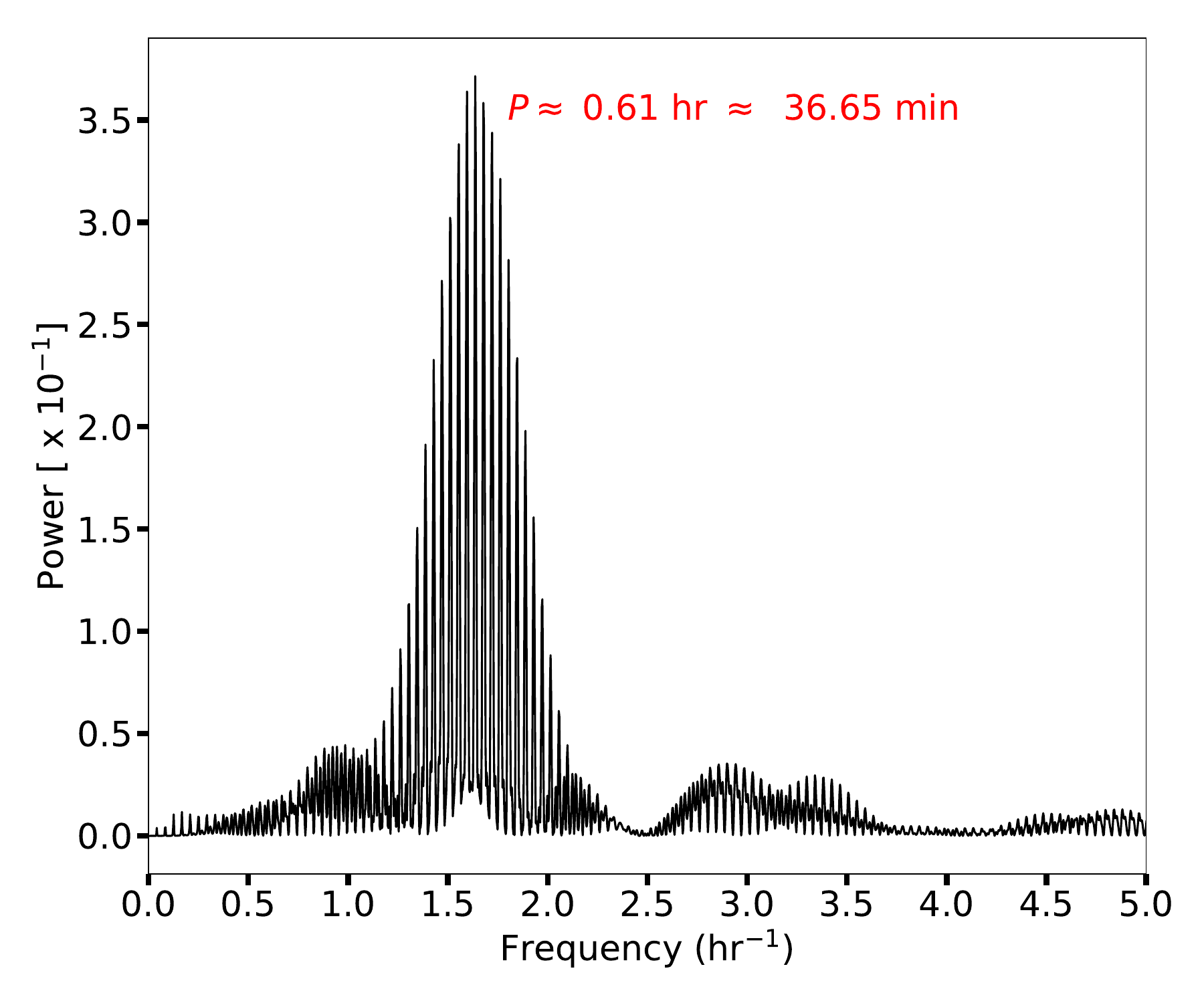}
        \includegraphics[width=0.49\textwidth]{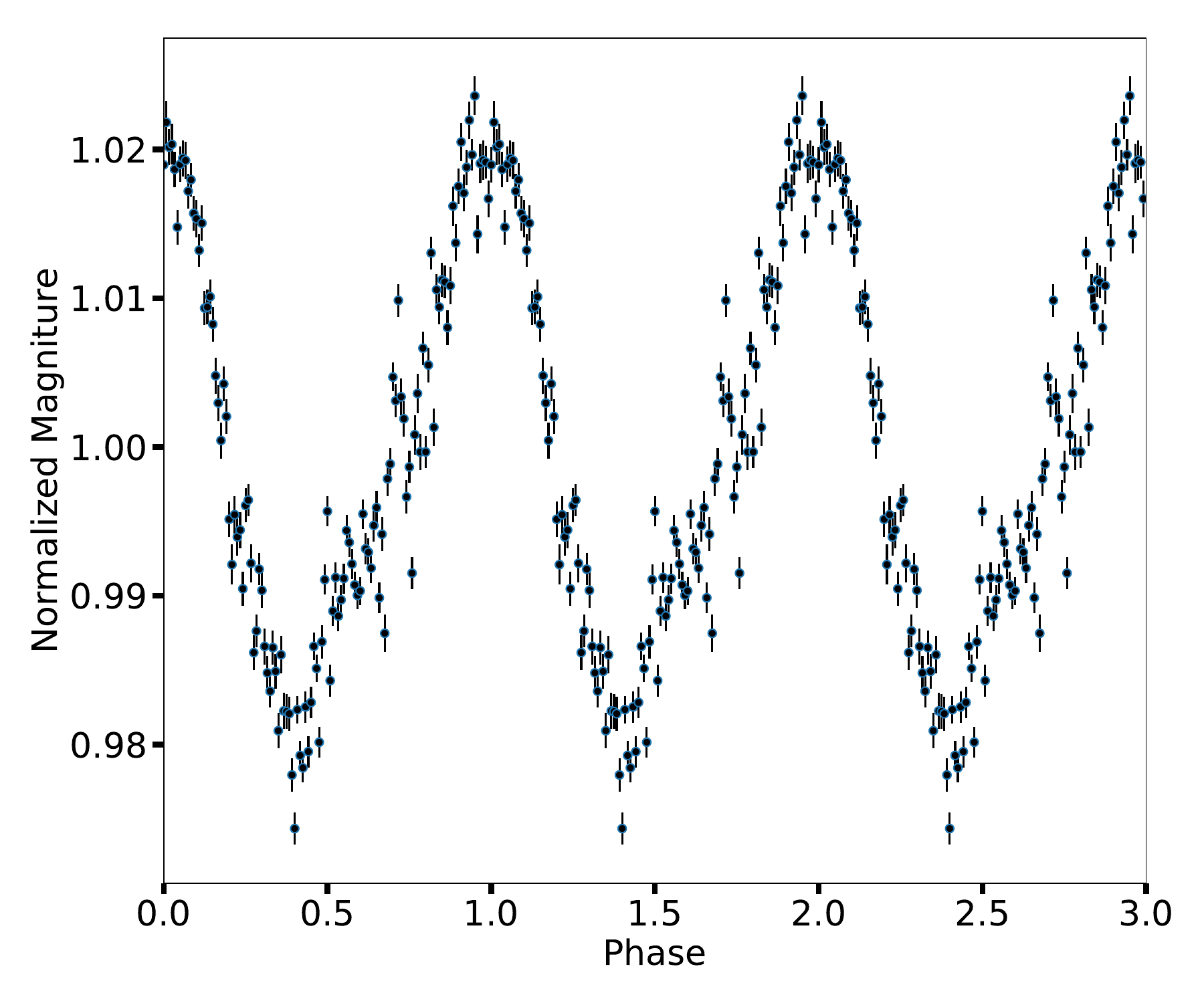}
    \caption{\textit{Left:} a Lomb-Scargle periodogram of the SHOC photometry. The periodogram shows a peak at a period of around 36.65 min. \textit{Right:} The phased SHOC data folded over the 36.65 min period.}
    \label{fig:Periodogram}
\end{figure*}

The behaviour of AM CVn systems vary depending on their orbital periods. The standard theory predicts that for systems with periods of less than 10 minutes, the accretion stream directly impacts onto the primary white dwarf, and no accretion disc is formed \citep{Solheim_2010}. For systems with  periods between around 10 and 20 min, the accretion disc is thermally stable in a hot state (the temperature of the disc is always greater than the ionization temperature of helium), and thus does not experience outbursts. Systems with orbital periods between around 20 and 44 minutes, have unstable accretion discs and experience outbursts where their brightness increases by a few mag for days to weeks \citep{Ramsay_etal_2012}. 
For systems with the longest orbital periods of 44 to 70 minutes, the accretion disc is in a thermally stable low state (the temperature of the disc is always lower than the ionization state of helium), and are expected not to have outbursts or to only rarely have one. However, the long period boundary below which outbursts are observed is not sharp, and many long period (between 50 and 60 minutes) systems were observed to show outbursts \citep{Ramsay_etal_2018,Rivera_Sandoval_etal_2021,Rivera_Sandoval_etal_2022,Maccarone_etal_2024}. Similarly, some short period systems were recently discovered to be in a stable high-state while having periods of 20 to 44 minutes \citep{Duffy_etal_2021}. This implies that the traditional classifications, which  seemed to agree well with the predictions from accretion disc instability models, are not necessarily fully representative of the behaviour of AM CVn systems; that is, with more systems discovered and their orbital periods determined, it is clear that our understanding of the behaviour of AM CVns is still fairly incomplete. Therefore, it is of a great value to increase the sample of known outbursting AM CVns with determined orbital periods to improve our understanding of the outburst mechanisms and constrain the physics of accretion disc models.

A typical outbursting AM CVn system increases in brightness by 2\,--\,5 magnitudes, over a few days up to a few weeks \citep{Ramsay_etal_2012,van_Roestel_etal_2021,Duffy_etal_2021}. The duration of the outbursts was first suggested to be correlated with the orbital period \citep{Ramsay_etal_2012}. However, \citet{Cannizzo_Ramsay_2019} and \cite{Duffy_etal_2021} later found no obvious correlation between these two parameters. This is primarily because other parameters, such as the mass of the primary white dwarf, play a role in determining the characteristics of the outbursts and their frequencies. Long term monitoring of AM CVns showed that some systems exhibit two types of outbursts: ``normal" outbursts, which last for only a few days with an amplitude of a couple magnitudes, while ``superoutbursts" can last more than 10 days with an amplitude of more than 3 magnitudes  \citep{Solheim_2010,Ramsay_etal_2012,Ramsay_etal_2018}. For some systems, a superoutburst can be preceded by a normal outburst \citep{Duffy_etal_2021,Marcano_etal_2021}.
During a superoutburst, periodic brightness fluctuations, known as superhump, occur with periods similar to the orbital period \citep{Patterson_etal_2005,Solheim_2010}. These are thought to be due to an interaction between the primary star and an eccentric precessing disc.

ASASSN-21br was discovered by the All-Sky Automated Survey for Supernovae (ASAS-SN; \citealt{Shappee_etal_2014,Kochanek_etal_2017}) on 2021-02-13.12UT as an optical transient at a discovery magnitude $g=12.32$. The transient is located at (J2000) equatorial coordinates of $(\alpha, \delta)$ = (16$^{\mathrm{h}}$18$^{\mathrm{m}}$10$^{\mathrm{s}}$\!\!\!.\,44,
$-$51$^{\circ}$54$^\prime$15$^{\prime\prime}$\!\!\!.\,8) and Galactic coordinates of ($l, b$) = (331$^{\circ}$\!\!\!.\,889, $-1{\circ}$\!\!\!.\,05). There is a Gaia DR3 (J2016)  \citep{Gaia_Collaboration_2023} source (ID 5934679711067619072) matching the transient position with an average $G$ magnitude of 19.7. Archival photometry from ASAS-SN showed no previous outbursts of the system dating back to March 2016. Spectroscopic follow up observations carried out by \citet{ATel_14405,ATel_14421} showed a spectrum dominated by emission lines of \eal{He}{I} and \eal{He}{II} and no hydrogen lines. \citet{ATel_14421} also found periodic modulations of $\approx$ 38 minutes, in photometric observations taken during the outburst. This led them to classify ASASSN-21br as an AM CVn system.

In this paper we present time-resolved photometric and spectroscopic follow-up observations of ASASSN-21br taken during its outburst to try to determine the orbital period of the system and explore the outburst properties. In Section~\ref{Obs} we present the observations and data reduction, while in Section~\ref{Results} we present the results of the spectral and photometric analysis. Further discussion and conclusions are presented in Section~\ref{Disc}.

\begin{figure*}
    \centering
    \includegraphics[width=\textwidth]{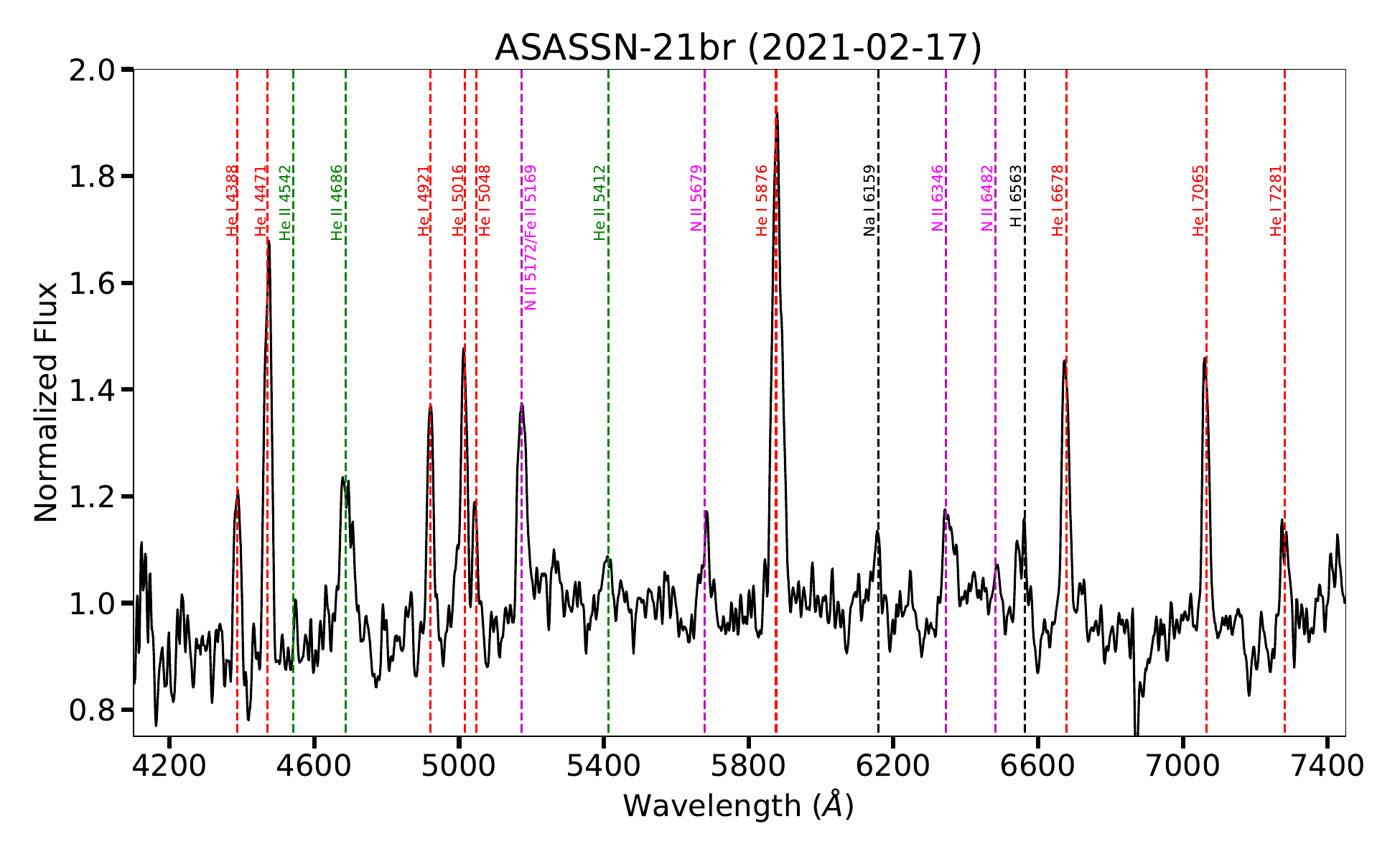}
    \caption{The SOAR spectrum of ASASSN-21br taken on 2021-02-18 during the dip in the light curve (first dashed line in Figure~\ref{fig:ASASSN_LC}).}
    \label{fig:SOAR spec}
\end{figure*}

\begin{figure*}
    \centering
    \includegraphics[width=\textwidth]{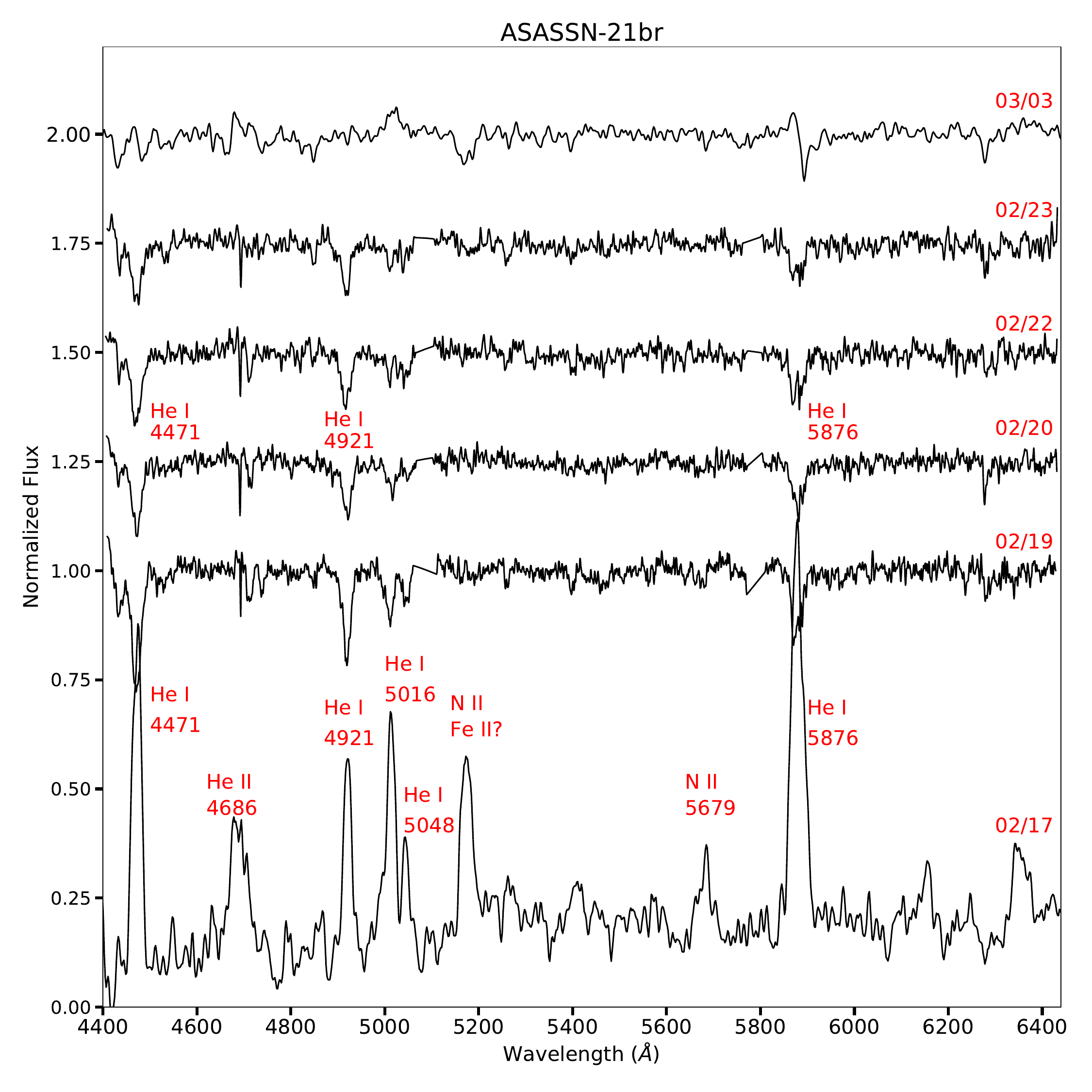}
    \caption{The spectral evolution of ASASSN-21br during its outburst. A vertical offset is added to the spectra for visualization purposes. Line identifications are included to guide the reader.}
    \label{fig:spectral evolution}
\end{figure*}

\begin{figure*}
    \centering
    \includegraphics[width=0.48\textwidth]{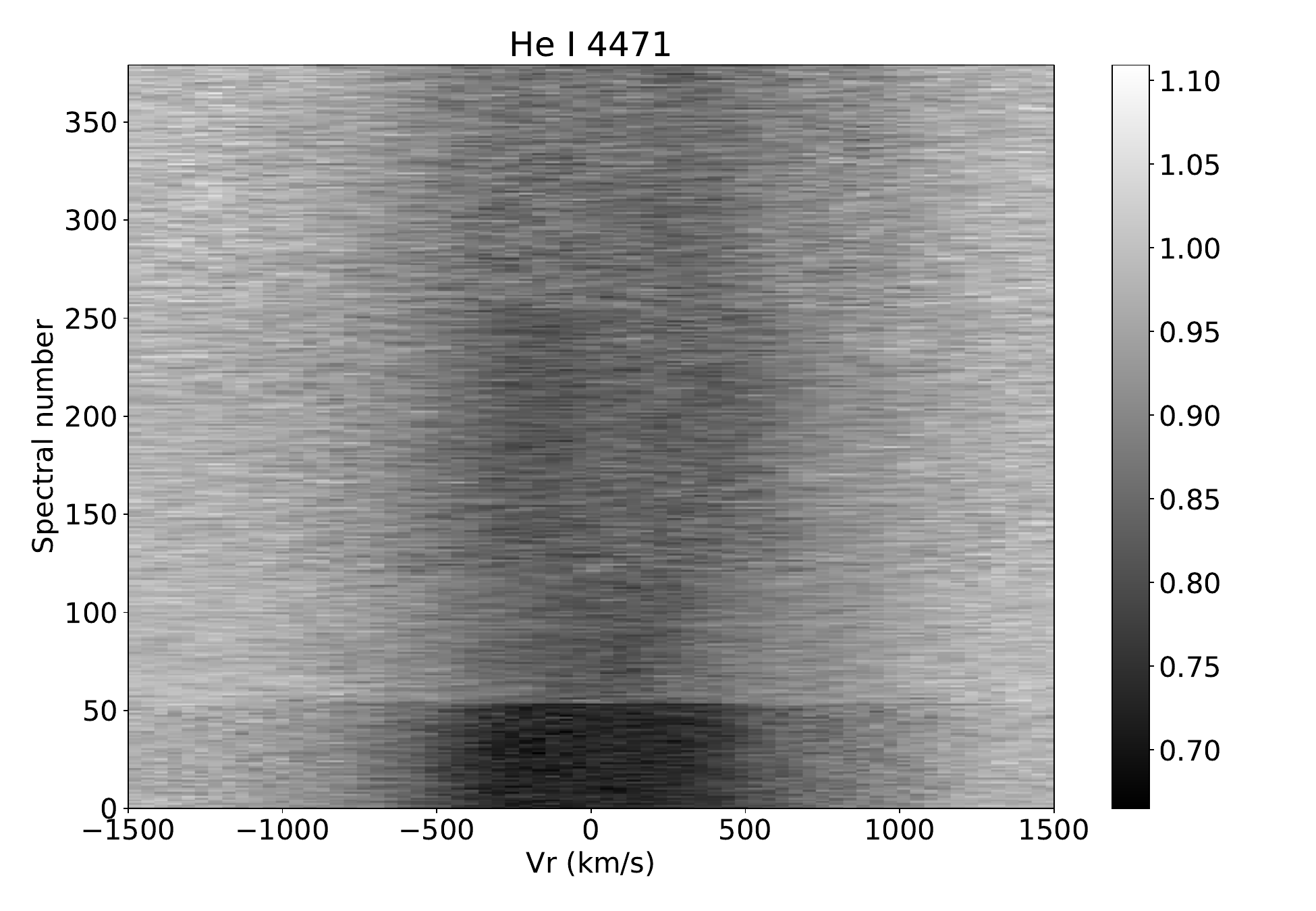}
    \includegraphics[width=0.48\textwidth]{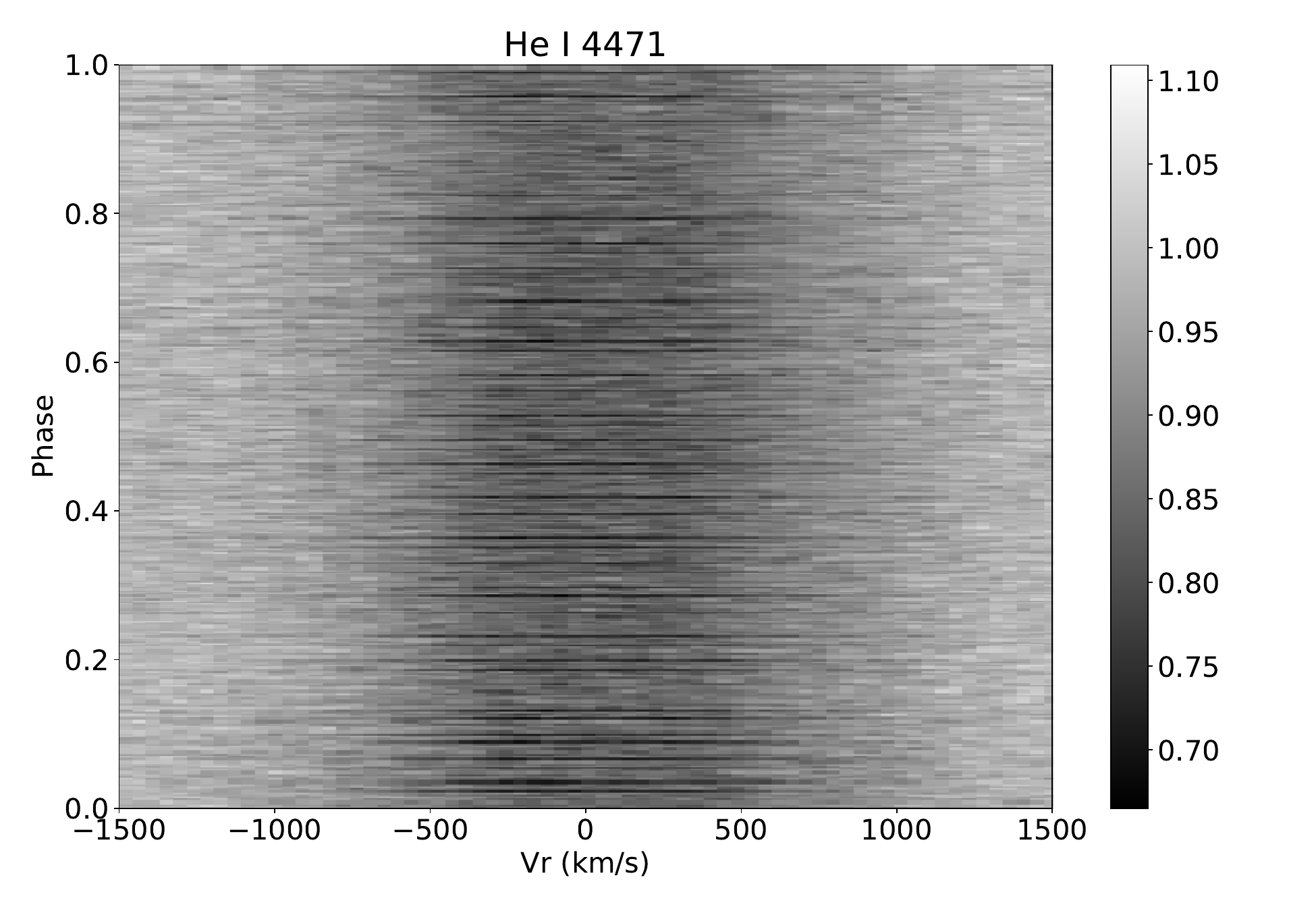}
    \caption{2D dynamic spectra of ASASSN-21br centered on the He I 4471\,\AA\,\,absorption line. The left panel shows the spectra in chronological order. The right panel shows the spectra sorted by phase based on the photometric period. A barycentric correction is applied to the radial velocities.}
    \label{Fig:dynamic_spec_HeI4471}
\end{figure*}

\begin{figure*}
    \centering
    \includegraphics[width=0.48\textwidth]{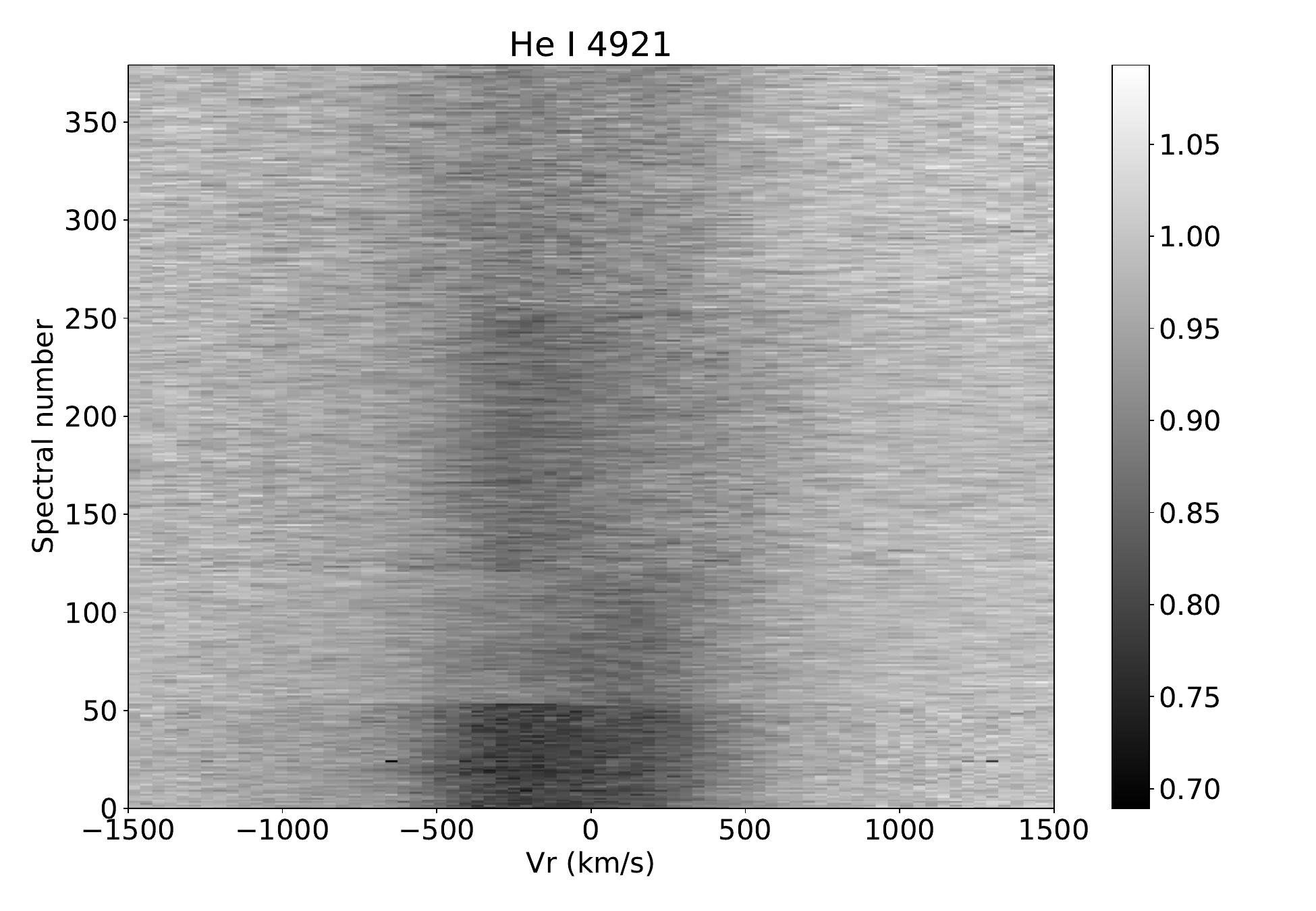}
    \includegraphics[width=0.48\textwidth]{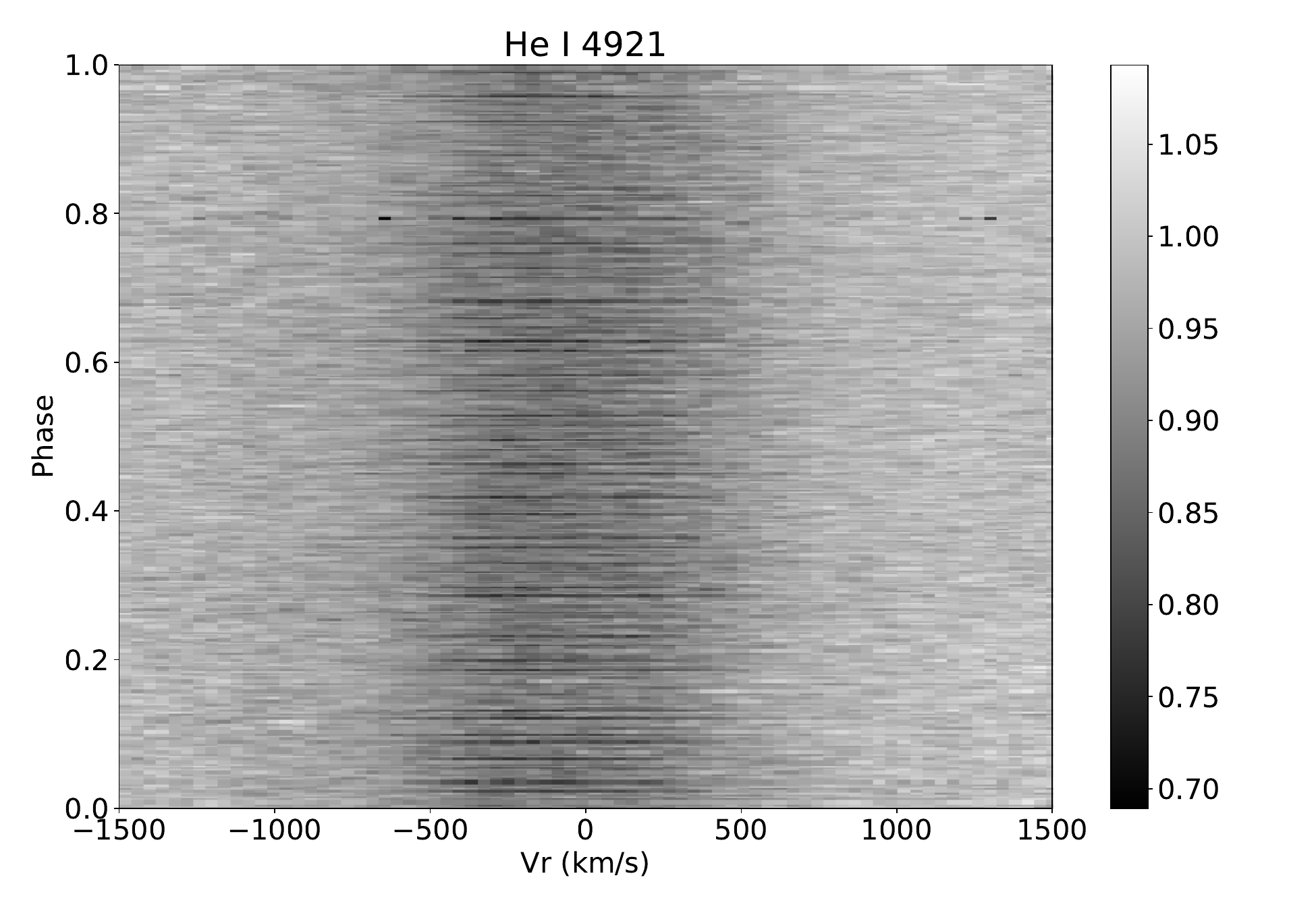}
    \caption{Same as Figure~\ref{Fig:dynamic_spec_HeI4471} but centered on the He I 4921\,\AA\,\,absorption line.}
    \label{Fig:dynamic_spec_HeI4921}
\end{figure*}

\section{Observations and Data Reduction}
\label{Obs}

ASSASN-21br was observed using the Sutherland High-speed Optical Camera (SHOC; \citealt{Gulbis_etal_2011,Coppejans_etal_2013}) mounted on the  1.0 meter telescope at the South African Astronomical Observatory (SAAO) in Sutherland, South Africa. The time resolved photometry was obtained on the nights of 2021 February 19th, 20th, 21st, and 22nd.
The 5 or 10\,s exposures were taken with a $g$' filter. The CCD images were flat-fielded and reduced using the \textsc{Image Reduction and Anaylsis Facility} (\textsc{IRAF}; \citealt{Tody_1986}) tasks. \textsc{PHOT} from the \textsc{DAOPHOT} package was used to perform aperture photometric reduction and \textsc{MKAPFILE} from the \textsc{PHOTCAL} package was used to apply an aperture correction. Comparison stars within the $2.85 \times 2.85$\,arcmin$^{2}$ frame were used to differentially correct the relative target magnitudes. 


We obtained time-resolved spectroscopy of ASASSN-21br on the nights of 2021 February 19th, 20th, 22nd, and 23rd using the Robert Stobie spectrograph (RSS; \citealt{Burgh_etal_2003,Kobulnicky_etal_2003}) mounted on the 10-m Southern African Large Telescope (SALT; \citealt{Buckley_etal_2006,Odonoghue_etal_2006}) in Sutherland, South Africa. We obtained a total of 388 spectra over the course of the four nights, each with a 60\,s exposure, using the PG1300 grating with a 1.5\,arcsec slit, resulting in a resolving power $R \approx 1500$ over the wavelength range 4400\,--\,6400\,\AA. Initial reduction was performed using the \textsc{PySALT} pipeline \cite{Crawford_etal_2010}, involving bias subtraction, cross-talk correction, scattered light
removal, bad pixel masking, and flat-fielding. Further reduction, including wavelength calibration, background subtraction, spectral extraction, flux calibration, and normalization was done using \textsc{IRAF}.

We also carried out low--resolution optical spectroscopy on the nights of 2021-02-17 and 03-03 using the Goodman spectrograph \citep{Clemens_etal_2004} mounted on the 4.1\,m Southern Astrophysical Research (SOAR) telescope located on Cerro Pach\'on in Chile. The observations were taken using the 400~l\,mm$^{-1}$ grating and a 1\arcsec\.2 slit, yielding a resolving power $R \approx$ 1100 over the wavelength range 3820--7850\,\AA. The spectra were reduced and optimally extracted using the \textsc{apall} package in \textsc{IRAF}. The spectroscopic observation log is presented in Table~\ref{table:spec_ASASSN_21br}.

We also make use of publicly available photometry from ASAS-SN \citep{Shappee_etal_2014}. These consist of $g$-band photometry taken as part of the ASAS-SN sky patrol survey \citep{Kochanek_etal_2017}.

\section{Results}
\label{Results}
\subsection{The outburst of ASASSN-21br}

Based on the ASAS-SN photometry, the outburst of ASASSN-21br started around 2021-02-13.12UT (HJD 2459258.62; see Figure~\ref{fig:ASASSN_LC}). It reached a peak brightness at $g \approx$ 13.5 and then gradually declined to around 13.8 in 2.5 days. The system then showed a sudden drop in brightness, dimming below the ASAS-SN sensitivity limit ($g \gtrsim 16.5$) between days 3 and 8, before showing a sudden recovery in brightness, reaching magnitudes brighter than 14\,mag. Such brightness dips during outbursts are common in AM CVn systems \citep{Ramsay_etal_2012,Ramsay_etal_2018,Duffy_etal_2021,Marcano_etal_2021}. Then, the brightness decreased over several days, before again dropping below the ASAS-SN sensitivity limit as the outburst ended. Searching the ASAS-SN archive, we do not find previous outbursts of ASASSN-21br, dating back to 2016-03-08. Similarly, the system has not had any additional outbursts.  

\subsection{Distance and absolute magnitudes}
The coordinates of the system are consistent with a Gaia DR3 source (ID 5934679711067619072; \citealt{Gaia_Collaboration_2023}), which has a mean magnitude of $G$ = 19.7 and a color of ($B_p-R_p$) = 0.8. The latter are extensively red compared to the colors of AM CVns during quiescence, however, the Gaia DR3 colors suffer from large uncertainties and therefore it is challenging to draw conclusions about the accretion rate or the disc/WD temperature based on the reported colors. The parallax of this source places ASASSN-21br at a distance of 515$^{+79}_{-54}$\,pc, based on the estimates from \citet{Bailer-Jones_etal_2021}. Since $g$ magnitudes are likely to be slightly dimmer than $G$ ones due to extinction (see below), we estimate that the amplitude of the outburst is $\gtrsim$ 6\,mag. This means that the outburst of ASASSN-21br is a superoutburst and is among the upper end of (super)outburst-amplitudes observed in AM CVn systems \citep{Solheim_2010,Ramsay_etal_2012,Ramsay_etal_2018,Rivera_Sandoval_etal_2021,Rivera_Sandoval_etal_2022}. 

We use the Gaia distance, the Galactic reddening maps of \citet{Chen_etal_2019}, and the extinction laws from \citet{Wang_etal_2019} to estimate the reddening towards ASASSN-21br. Based on the maps from \citet{Chen_etal_2019} we derive $E(G-K_s) = 0.170 \pm 0.007$, $E(G_{Bp}-G_{Rp})_s = 0.067 \pm 0.016$, and $E(H-K_s) = 0.0045 \pm 0.0005$. This translates to $A_V = 0.16 \pm 0.04$, $E(B-V) = 0.05 \pm 0.02$, $A_G = 0.12 \pm 0.05$, and $A_g= 0.19 \pm 0.05$. 
With these reddening values, we derive a peak absolute magnitude of $M_g = 4.8 \pm 0.4$ and an absolute magnitude at quiescence of $M_G = 11.02 \pm 0.4$. These are both within the typical range of brightness during outburst and quiescence for the orbital period of the system ( \citealt{Ramsay_etal_2012,Levitan_etal_2015}; see sections below). 

The system is detected by \textit{Chandra} during a 40\,ks observations of the planetary nebula MZ 3 taken on 2002-10-23. While ASASSN-21br is 10' off-axis in the \textit{Chandra} observations, the \textit{Chandra} source catalogue \citep{2010ApJS..189...37E} flux of the system is around $5.8 \times 10^{-14}$\,erg\,s$^{-1}$\,cm$^{-2}$ (source ID 2CXO J161810.3-515414). This means that the X-ray luminosity of the system during quiescence is around $1.8 \times 10^{30}$\,erg\,s$^{-1}$. This is consistent with the X-ray luminosities of AM CVn systems during quiescence \citep{Ramsay_etal_2005,Ramsay_etal_2006,Begari_Maccarone_2023}.

\subsection{Photometric timing analysis}

We performed timing analysis for the 4 nights of SHOC photometry searching for periodicity. We show a sample of the photometric data from one of these nights (2021-02-21) in Figure~\ref{fig:Photometry}. We computed the Lomb-Scargle periodogram \citep{Lomb_1976,Scargle_1982,2018ApJS..236...16V} after detrending the light curve for (1) each night separately and (2) all the nights combined. The periodograms of each separate night show periodic modulation between around 35 and 38 minutes (Figure~\ref{Fig:LSP_all}). The periodogram of the combined data shows modulation at a period of $P= 0.025448 \pm 0.000106$\,days ($\approx$ 36.65 mins; Figure~\ref{fig:Periodogram}). The period uncertainty is conservatively estimated as $P_{\rm error} = P^2/(2 T)$, 
corresponding to the period change that would cause the phase shift of 0.5 between the first and the last points in the light curve. The phase folded, binned light curve is plotted in Figure~\ref{fig:Periodogram}, showing a nearly sinusoidal trend. The 36.65 min period is within the range of orbital periods of outbursting AM CVn systems, which are typically characterized by a $P_{\mathrm{orb}}$ between 20 and 44 minutes \citep{Ramsay_etal_2012}. However, given that the photometry are taken during outburst, the period could also be the superhump period \citep{Patterson_etal_2005,Solheim_2010}. The change of a few minutes in the period modulation between our different epochs is not uncommon for superhump periods during  outbursts of AM CVns and dwarf novae \citep{Isogai_etal_2016,Kato_etal_2016_PASJ...68...65K}.

\subsection{Spectral evolution}

The first spectrum of ASASSN-21br (Figure~\ref{fig:SOAR spec}), taken on 2021-02-18 during the dip in the light curve (around 5 days since the eruption start), was dominated by strong emission lines of He I, He II, and N II. The Full Width at Half Maximum (FWHM) of the He I lines was around 1300\,---\,1600\,km\,s$^{-1}$, which is within the typical range of velocities observed in AM CVns during outburst \citep{Anderson_etal_2005}. At the end of the dip, the spectra showed dramatic changes, switching from an emission line dominated spectrum to an absorption line dominated spectrum. The He II and N II lines almost completely disappear, with the spectrum showing mostly He I absorption lines. The FWHM of the He I absorption lines was also around 1300\,--\,1600\,km\,s$^{-1}$. The spectral evolution of ASASSN-21br 
throughout the outburst is shown in Figure~\ref{fig:spectral evolution}. The last spectrum obtained on 2021-03-03 still shows He I absorption lines, indicating that the outburst lasted longer than 20 days. We expect that by the end of the outburst, the spectrum would show emission lines again.

\subsection{Spectroscopic timing analysis}
We attempt to retrieve the photometric period from the time-resolved spectroscopic observations performed with SALT over four nights. We derive radial velocities from the strongest He I absorption lines at 4471, 4921, and 5876\,\AA\,\,using two methods: (1) we use Gaussian fits to derive the line centers and then calculate the radial velocities. This has been done using the \textsc{splot} package in \textsc{IRAF} and using the \textsc{Scipy} Python package; (2) we also computed radial velocities using \textsc{fxcor} in \textsc{IRAF} to do a Fourier cross-correlation. A high SNR spectrum is used as a template to derive the relative radial velocity shifts of each spectrum to the high SNR spectrum. This was done separately for each night. In both methods, we apply a barycentric correction to the radial velocities. The uncertainties are derived by combining (1) the average shift of the observed sky lines from their central wavelength and (2) the average shift of the strongest emission lines in the arc spectra used to wavelength calibrate the science spectra ($\Delta RV = \sqrt{(\Delta RV_{\mathrm{Sky}})^2 + (\Delta RV_{\mathrm{arc}})^2}$). In both cases, the shifts were derived by fitting a Gaussian to the lines. This resulted in uncertainties of a few 10s of km\,s$^{-1}$.

The radial velocities of the He I absorption lines show random changes and no periodic modulation. We show the radial velocity curves folded over the 36.65 min period in Figure~\ref{Fig:RV_curve}. A barycentric correction is applied to the observation dates (assumed as middle of the exposure). The radial velocities do not show any modulation consistent with the photometric period. In Figures~\ref{Fig:dynamic_spec_HeI4471} and~\ref{Fig:dynamic_spec_HeI4921} we present 2D dynamic trailed spectral evolution over the 4 nights centered on the He I 4471 and 4921\,\AA\,\,absorption lines, respectively. The 2D dynamic trailed spectra are presented in both chronological order and phase folded using the photometric period. The latter do not show any sinusoidal `S-shape' trend, again implying  that the changes in the spectral lines are not modulated over the photometric period.

\section{Discussion and Conclusions}
\label{Disc}

\subsection{The orbital period of ASASSN-21br}
The photometric period of 36.65 min, which we found in the SHOC data, is possibly representative of the orbital period of ASASSN-21br, given that it falls within the typical range of orbital periods of oubtursting AM CVn systems. However, since the data are taken during outburst, the period is highly likely the superhump period \citep{Patterson_etal_2005,Solheim_2010}. The difference between the orbital period and the superhump period is likely small (a few per~cent). The time-resolved spectroscopy could have helped us improve the orbital period estimates, but we did not find any periodic modulation in the spectral absorption lines taken during the outburst. We suggest that this is mostly due to the He I absorption lines originating in the outbursting accretion disc, and therefore are not ideal to determine the orbital period of the system. In order to get better constraints on the orbital period, time-resolved spectroscopy taken during quiescence is needed. This is both challenging and time-consuming task given the brightness of the system in quiescence (around 20 mag in the optical) and the broad emission lines in the spectrum originating from the accretion disc.

\subsection{The properties of the outburst}
The 2021 outburst of ASASSN-21br is the only outburst   recorded in the system between 2016 and 2024. While AM CVn systems experience multiple outbursts, the frequency of these outbursts might decrease with increasing orbital period \citep{Levitan_etal_2015,Ramsay_etal_2018,Duffy_etal_2021,Rivera_Sandoval_etal_2022}. \citet{Levitan_etal_2015} derived an empirical relation for the recurrence period as a function of the orbital period. Based on these relations, we derive a recurrence period of ASASSN-21br: $$P_{\mathrm{recur}} = (1.53 \times 10^{-9}) P_{\mathrm{orb}}^{7.35} + 24.7 \approx 504\,\, \mathrm{days}.$$

An accurate estimate for the recurrence period of ASASSN-21br is not available due to lack of coverage. ASAS-SN did not catch any other outburst during 1800 days (between the first ASAS-SN observation of the field of ASASSN-21br on 2016-03-18 and the start of the 2021 outburst; see Figure~\ref{Fig:full_LC}). While outbursts might have been missed due to solar constraints (see below), this is almost 4 times the estimate from the empirical relations of \citet{Levitan_etal_2015}. This shows that the behaviour of outbursting AM CVn systems is more diverse than previously thought \citep{Duffy_etal_2021,Rivera_Sandoval_etal_2021}.

\begin{figure}
    \centering
    \includegraphics[width=\columnwidth]{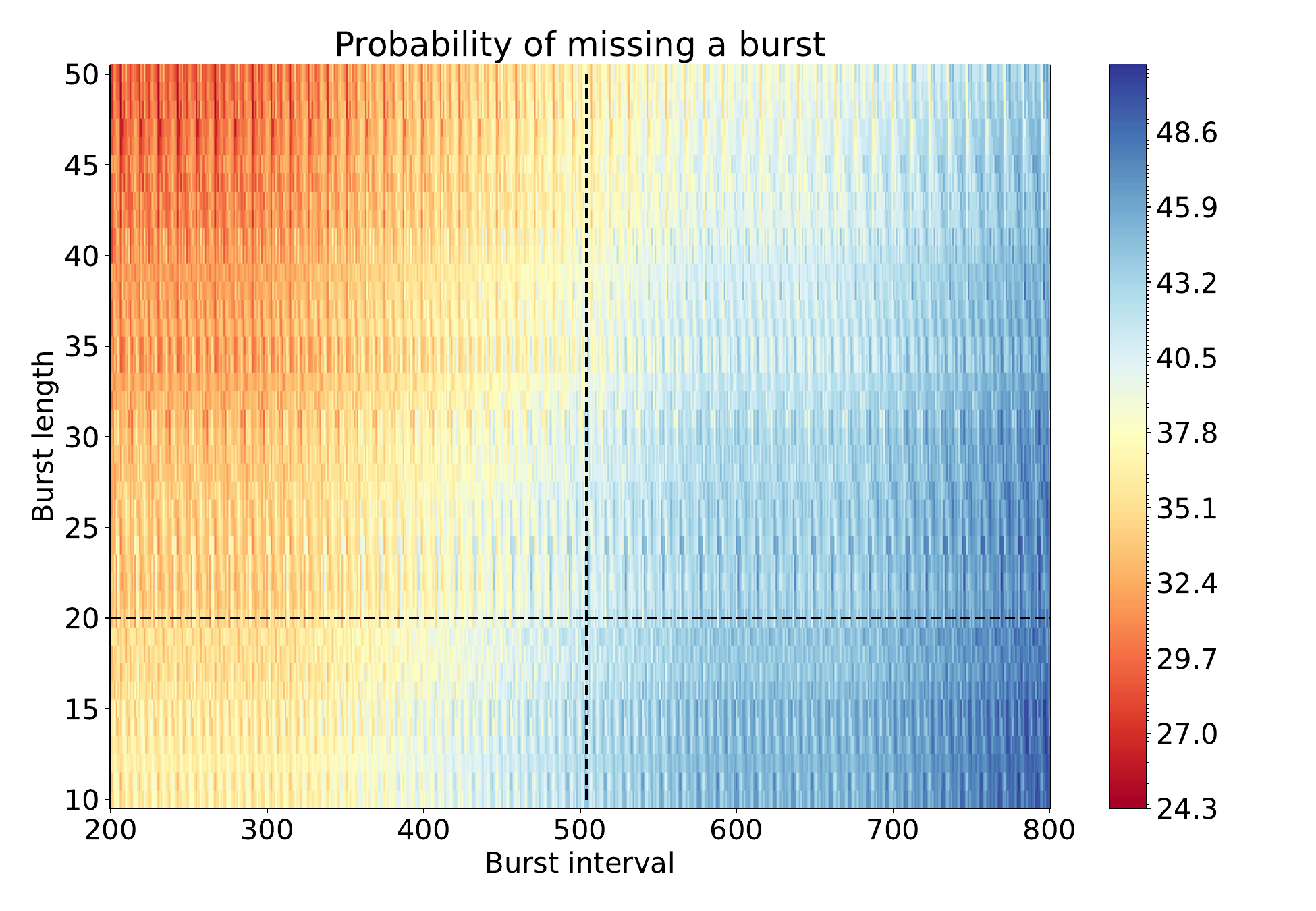}
    \caption{We show the probability (percentage) of missing an outburst as a function of outburst interval and outburst length. The horizontal dashed line represents the duration of the 2021 outburst of ASASSN-21br and the horizontal vertical line represents the inter-bust intervals as derived for ASASSSN-21br using the empirical relations of \citet{Levitan_etal_2015}; see text for more details.}
    \label{Fig:LC_simulated}
\end{figure}

The ASAS-SN coverage is interrupted by solar conjunction for around 3 months every year year. Therefore, there is a substantial chance of missing previous outbursts in the ASAS-SN data. In order to estimate the probability of missing one or more outbursts, we simulate $\approx$ 6$\times 10^6$ light curves for different burst lengths and inter-burst intervals, and compare these light curves with the ASAS-SN data (Figure~\ref{Fig:LC_simulated}). We found that there is a $\approx$ 25\% to 50\% chance of missing at least one outburst in the ASAS-SN data, with the chance decreasing with increasing burst length and decreasing inter-burst intervals. This is consistent with similar simulations performed by \citet{Ramsay_etal_2012}. For a burst interval of 504 days and a burst length of 20 days (as derived for ASASSN-21br using the relations of \citealt{Levitan_etal_2015}), the chance of missing an outburst is close to 45\%. This means that there is a significant possibility that a previous burst has been missed.

\begin{figure}
    \centering
    \includegraphics[width=\columnwidth]{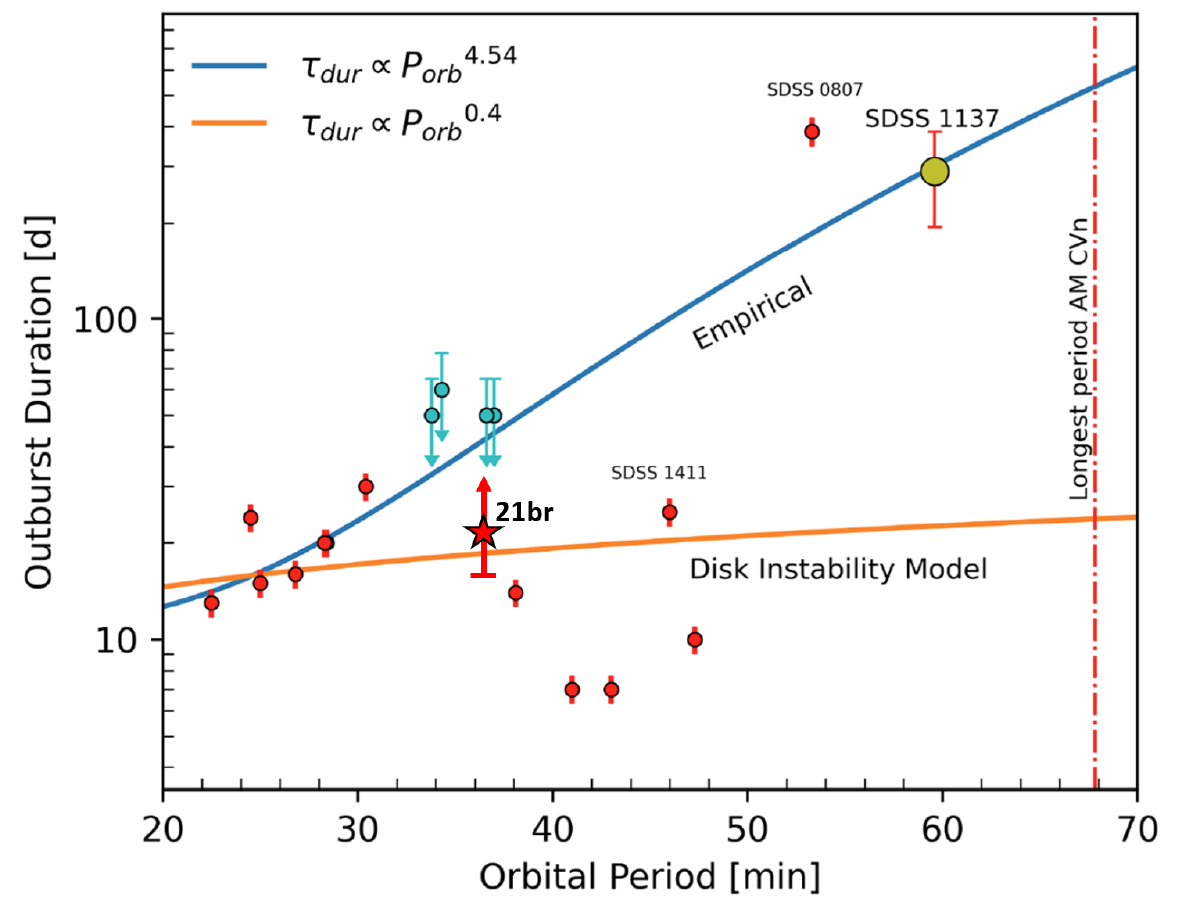}
    \caption{The outburst duration as a function of orbital period for several AM CVn systems, adopted from \citet{Rivera_Sandoval_etal_2021}, with ASASSN-21br as a red star. The blue line is the empirical relations of \citet{Levitan_etal_2015}, while the orange line is the relation from \citet{Cannizzo_Ramsay_2019} for the disc instability model.}
    \label{Fig:Porbvsdur}
\end{figure}

\citet{Levitan_etal_2015} also derived empirical relations relating the orbital period with the (super)outburst duration. It is challenging to determine the duration of the outburst of ASASSN-21br based on the ASASN-SN data alone, since the brightness of the system drops below ASASN-SN's sensitivity before the outburst is necessarily over. However, based on the ASAS-SN light curve and our spectroscopic follow up, we estimate an outburst duration of more than 20 days. In Figure~\ref{Fig:Porbvsdur} we compare the outburst properties of ASASSN-21br with other known AM CVn systems, and the empirical relations of \citet{Levitan_etal_2015}. The plot shows that the outburst duration of ASASSN-21br is not consistent with the relation of \citet{Levitan_etal_2015}, however it is consistent with disc instability models. This is true for the lower limit derived on the duration of the outburst. If the value is actually longer, the outburst duration would then be more consistent with the empirical relations of \citet{Levitan_etal_2015}.

\begin{figure}
    \centering
    \includegraphics[width=\columnwidth]{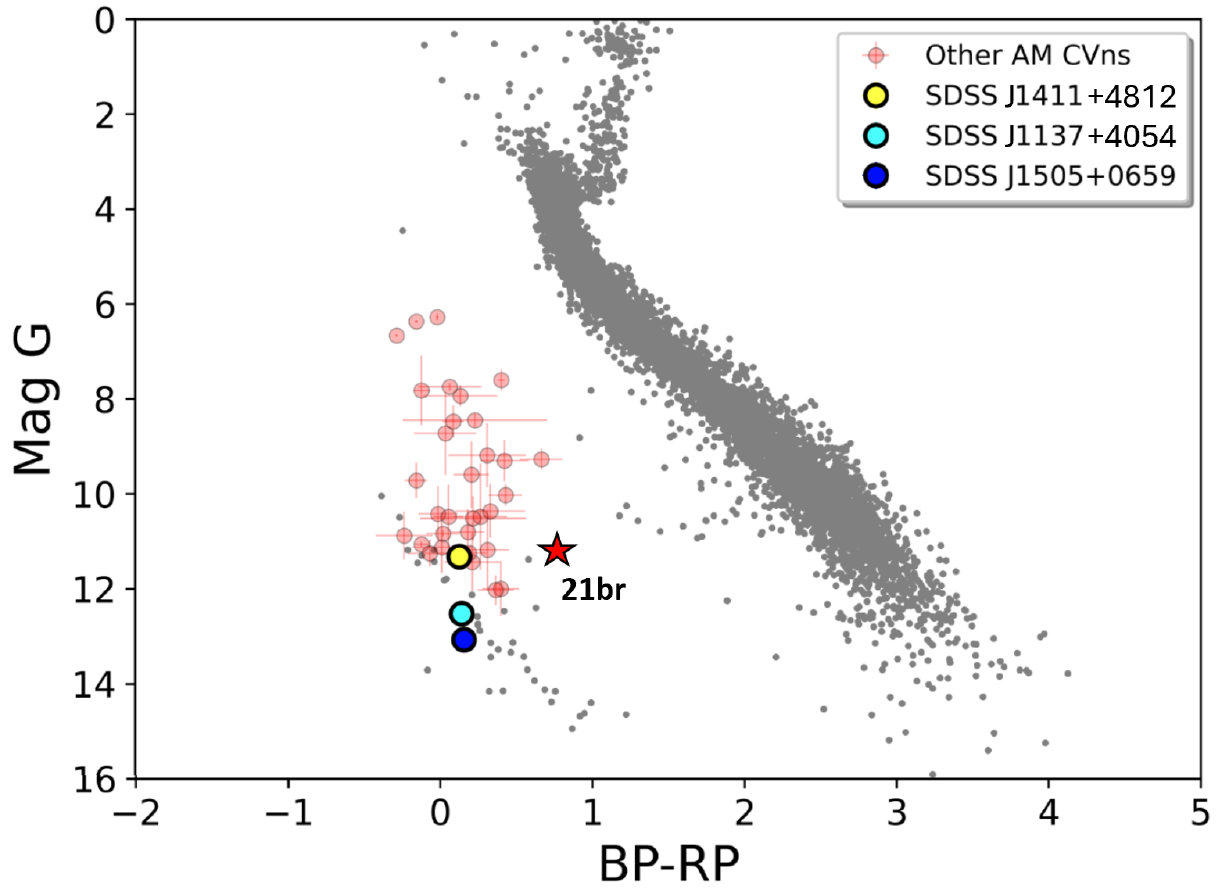}
    \caption{A Gaia color-magnitude-diagram of AM CVn systems and normal stars, adopted from \citet{Rivera_Sandoval_etal_2021}, with ASASSN-21br as a red star. The grey points are Galactic stars (near SDSS J113732+405458) with accurate Gaia parallaxes. No extinction correction is applied. The difference in the extinction between the direction of ASASSN-21br and SDSS J113732+405458 is small.} 
    \label{Fig:color_mag_diagram}
\end{figure}

In Figure~\ref{Fig:color_mag_diagram} we show a colour magnitude diagram, comparing the quiescent colours and absolute magnitude of ASASSN-21br to other AM CVn systems and other Galactic stars with accurate parallaxes (error of less than 10 per~cent). Like other AM CVn systems, ASASSN-21br is more luminous in the G-band then single white dwarfs of similar colour. This is mostly due to excess emission from the accretion processes. However, the system is fainter than several other AM CVn systems with shorter orbital periods---these are systems that are constantly in a high-state \citep{Ramsay_etal_2018}. Since the brightness of the system is suggested to be correlated with the orbital period \citep{Rivera_Sandoval_etal_2021}, and ASASSN-21br orbital period is on the longer end of outbursting AM CVns, we expect that the system is relatively dimmer during quiescence. This is also true when comparing ASASSN-21br to systems like SDSS J113732+405458 and SDSS J1505+0659, which are characterized by longer orbital periods ($P_{\mathrm{orb}}$ = 60\,min and 68\,min, respectively) and therefore are dimmer than ASASSN-21br in quiescence (Figure~\ref{Fig:color_mag_diagram}).

\subsection{The in-outburst dip}
Several AM CVn systems show pronounced dips during their (super)outbursts \citep{Ramsay_etal_2010,Levitan_etal_2011,Ramsay_etal_2012,Ramsay_etal_2018,van_Roestel_etal_2021,Duffy_etal_2021}. These dips are usually seen a few days after peak brightness and last for only a few days before the system brightens again. The origin of these dips is poorly understood, and whether they are similar to the dips observed in the WZ Sge-type hydrogen-accreting, dwarf novae \citep{Kotko_etal_2012}, or if they are related to the ``cycling state"---a series of low-amplitude, frequent outbursts, which prolong the duration of the superoutburst and may last as long as the smooth superoutburst itself \citep{Patterson_etal_1997,Patterson_etal_2000}---is not clear. For hydrogen-accreting cataclysmic variables, the dips are proposed to be a normal outburst
triggering a superoutburst \citep{van_der_Woerd_van_Paradijs_1987,Kato_1997}. However, this is challenged by the fact that the brightness of the first peak is larger than that of the second peak.  \citet{Kotko_etal_2012} attempted to reproduce the dips in AM CVn outbursts by changing the mass-transfer rate after the first peak. The enhanced mass-transfer rate leads to dips and echo outbursts.  

The first spectrum we obtained for ASASSN-21br was taken during the dip. The spectrum was emission lines dominated with a flat continuum (Figure~\ref{fig:SOAR spec}), before turning into an absorption line dominated spectrum after the dip (see Figures~\ref{fig:ASASSN_LC} and~\ref{fig:spectral evolution}). The shift from an emission to an absorption spectrum means that the disc switched states from optically thin to optically thick. 
During the dip, the spectrum is typical of spectra taken in quiescence, while after the dip, the spectrum is typical of outburst spectra.  
If the spectrum during peak (before the dip) was also absorption line dominated, this might imply that the dip is a phase between two outbursts, i.e., the system returned to a quiescent phase for a short period before another outburst was triggered. However, we do not have any spectra covering the pre-dip peak. 

\subsection{Summary}
In summary, we obtained photometric and spectroscopic follow up of the Galactic transient ASASSN-21br. The hydrogen-deficient spectra are consistent with that of an AM CVn system. Time-resolved photometry showed modulation with a period of 36.65 min, which is consistent with the period range of AM CVns. Given that these photometric observations were taken during the outburst, it is possible that they might represent the superhump period, which would be expected to be few per~cent larger than the actual orbital period. We attempted to retrieve the photometric period using time-resolved spectroscopy, also obtained during outburst, but we failed to find any periodicity possibly because the spectral absorption lines used to derive the radial velocities originate in the outbursting accretion disc. There is a pronounced dip in the light curve a few days after peak brightness, which is common in AM CVns. Our spectra show a significant change between the dip (emission lines) and post-dip (absorption lines), indicating a major change in the state of the disc during the dip. 

Increasing the sample of known AM CVn systems with derived orbital periods is important for future gravitational wave missions, and to improve our understanding of the outburst mechanism in these compact binaries. Future spectroscopy of ASASSN-21br in quiescence would aid in an improved constraints on its orbital period.

\section*{Acknowledgements}

S.P. acknowledges NASA award 80NSSC23K0497.
E.A. acknowledges support by NASA through the NASA Hubble Fellowship grant HST-HF2-51501.001-A awarded by the Space Telescope Science Institute, which is operated by the Association of Universities for Research in Astronomy, Inc., for NASA, under contract NAS5-26555.  JS was supported by the Packard Foundation. MM and DAHB gratefully acknowledges the receipt of research grants from the National Research Foundation (NRF) of South Africa. 

A part of this work is based on observations made with the Southern African Large Telescope (SALT), with the Large Science Programme on transients 2021-2-LSP-001 (PI: DAHB). Polish participation in SALT is funded by grant No. MEiN 2021/WK/01. This paper was partially based on observations obtained at the Southern Astrophysical Research (SOAR) telescope, which is a joint project of the Minist\'{e}rio da Ci\^{e}ncia, Tecnologia e Inova\c{c}\~{o}es (MCTI/LNA) do Brasil, the US National Science Foundation's NOIRLab, the University of North Carolina at Chapel Hill (UNC), and Michigan State University (MSU).

We thank Las Cumbres Observatory and its staff for their continued support of ASAS-SN. ASAS-SN is funded in part by the Gordon and Betty Moore Foundation through grants GBMF5490 and GBMF10501 to the Ohio State University, and also funded in part by the Alfred P. Sloan Foundation grant G-2021-14192. Development of ASAS-SN has been supported by the NSF, the Mt. Cuba Astronomical Foundation, the Center for Cosmology and AstroParticle Physics at the Ohio State University, the Chinese Academy of Sciences South America Center for Astronomy (CAS-SACA), and the Villum Foundation.

Analysis made significant use of \textsc{python} 3.7.4, and the associated packages \textsc{numpy}, \textsc{matplotlib}, \textsc{scipy}. 

Data reduction made significant use of \textsc{PySALT} \citep{Crawford_etal_2010}, and \textsc{IRAF} \citep{Tody_1986,Tody_1993}.

\section*{Data availability}
The data are available as online material and can be found here: \url{https://www.dropbox.com/scl/fi/ybbmjjbofu87zl60ulwww/online_material.zip?rlkey=qw4aeuo4879q03uq3kjmllahm&st=109qph6u&dl=0}.\\

\bibliographystyle{mnras_vanHack}
\bibliography{biblio}

\clearpage
\appendix

\renewcommand\thetable{\thesection.\arabic{table}}    
\renewcommand\thefigure{\thesection.\arabic{figure}}   
\setcounter{figure}{0}

\section{Supplementary plots}
\label{appB}
In this Appendix we present supplementary plots.

\begin{table}
\centering
\caption{Log of the spectroscopic observations of ASASNN-21br.}
\begin{tabular}{lccc}
\hline
Date & Instrument & $R$ & Range ($\mathrm{\AA}$)\\
\hline
\hline
2021-02-17 & SOAR-Goodman & 1000 & 4100\,--\,7500\\
2021-02-19 & SALT-RSS & 800 & 4400\,--\,6400\\
2021-02-20 & SALT-RSS & 800 & 4400\,--\,6400\\
2021-02-22 & SALT-RSS & 800 & 4400\,--\,6400\\
2021-02-23 & SALT-RSS & 800 & 4400\,--\,6400\\
2021-03-03 & SOAR-Goodman & 1000 & 4100\,--\,7500\\
\hline
\end{tabular}
\label{table:spec_ASASSN_21br}
\end{table}
\begin{figure*}
    \centering
    \includegraphics[width=\textwidth]{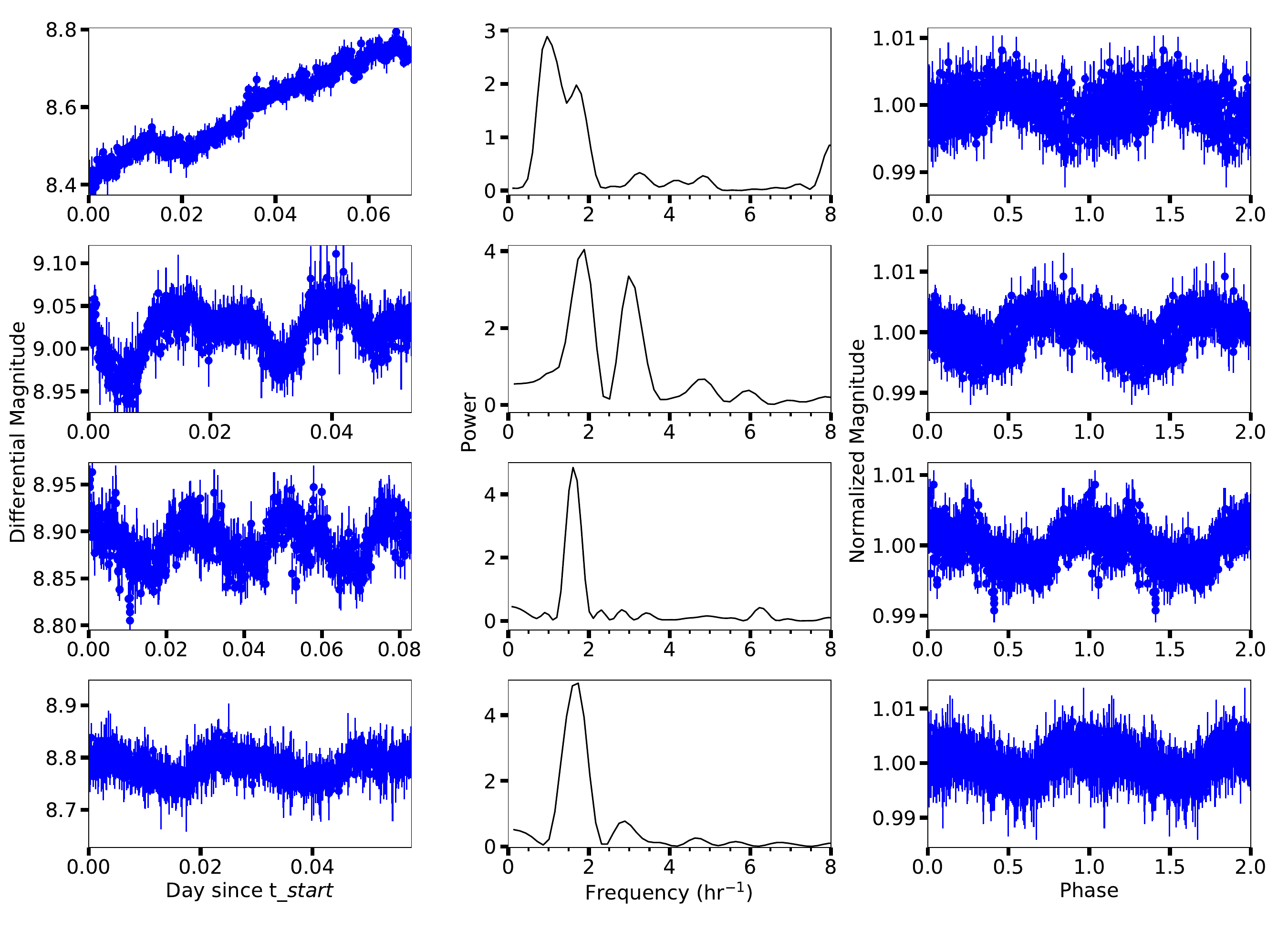}
    \caption{\textit{Left}: the SHOCH photometric data taken on the nights of 2021-02-19, 20, 21, and 22 (from top to bottom). \textit{Middle}: the Lomb-Scargle periodogram performed for each night. A detrending has been applied to the photometry prior to the timing analysis. \textit{Right}: the phased light curves folded over the period derived from the timing periodograms.}
    \label{Fig:LSP_all}
\end{figure*}

\begin{figure*}
    \centering
    \includegraphics[width=\textwidth]{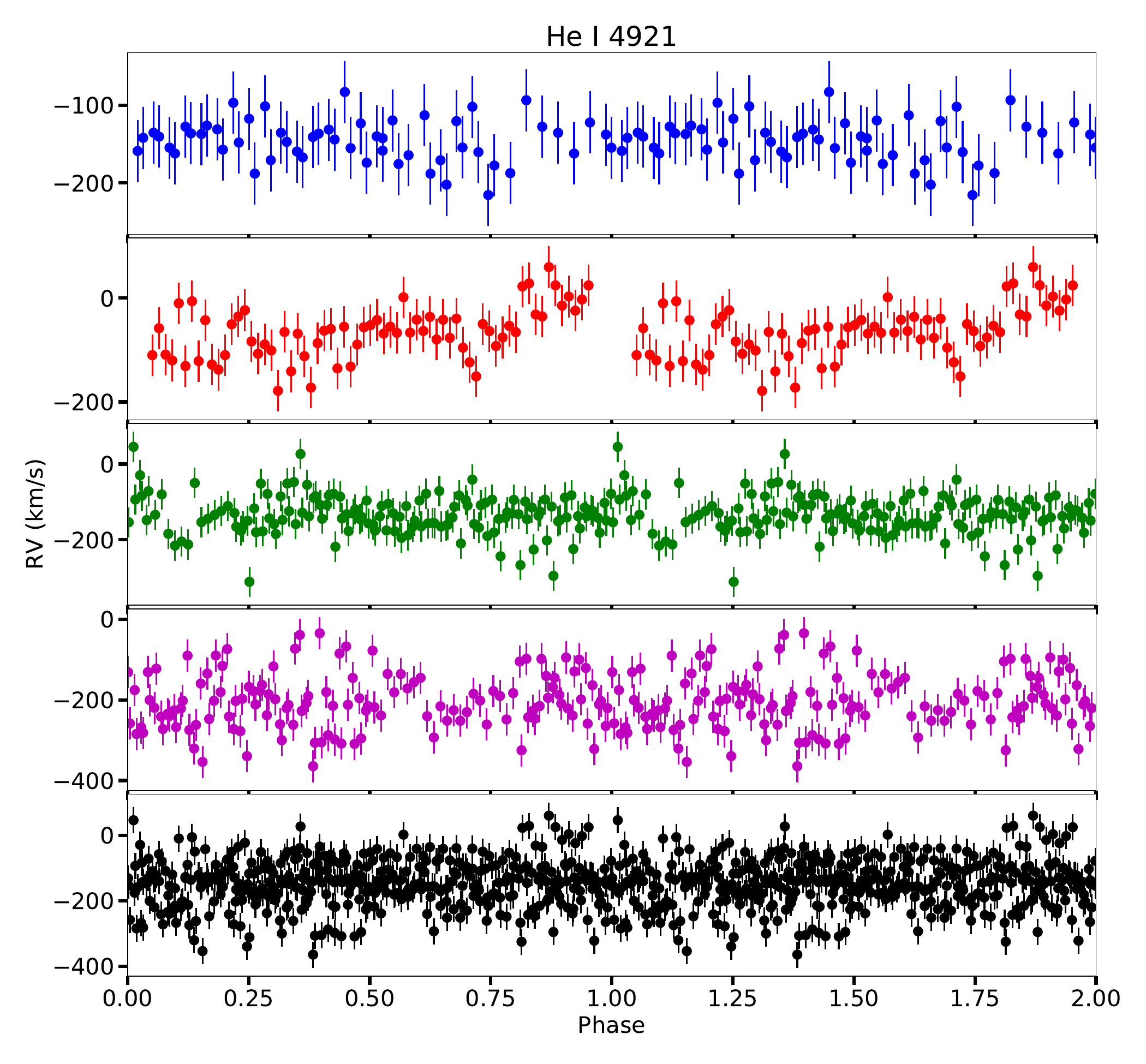}
    \caption{Radial velocity curves of ASASSN-21br folded over the 36.5 min period. From bottom to top, the nights of 19, 20, 22, and 23 of February 2021. The bottom panel shows the  measurements from all four nights. The barycentric -corrected velocities are measured using the He I 4921\,\AA\,\, absorption line.}
    \label{Fig:RV_curve}
\end{figure*}

\begin{figure*}
    \centering
    \includegraphics[width=\textwidth]{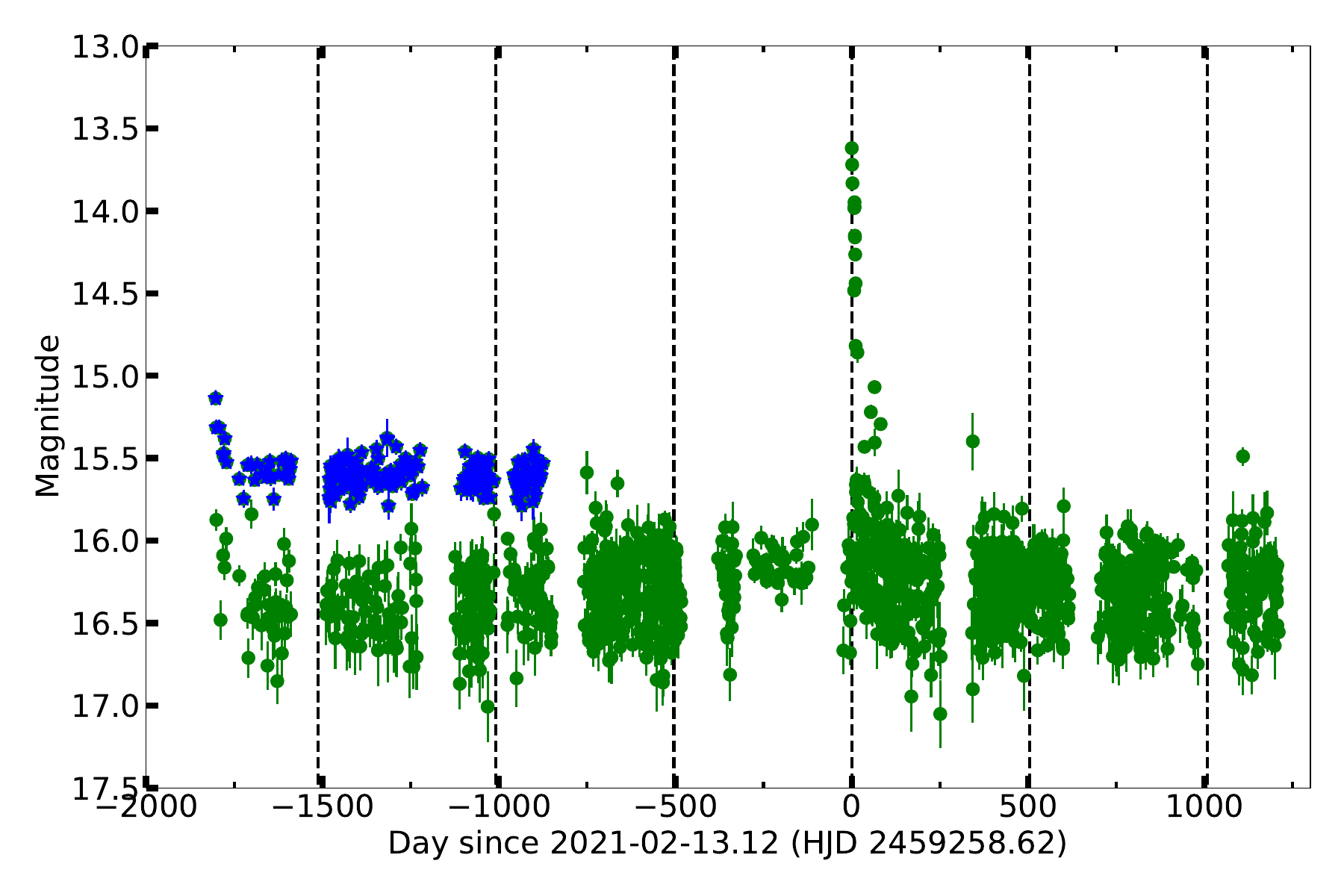}
    \caption{The ASAS-SN light curve of ASASSN-21br between 2016-03-08 and 2024-06-07. The green circles are $g$-band magnitudes while the blue stars are $V$-band magnitudes. The quiescent flux is dominated by nearby sources. The vertical dashed lines represent a 504 days duration centered at the 2021 superoutburst, which is considered as $t=0$.}
    \label{Fig:full_LC}
\end{figure*}

\end{document}